\documentstyle[aps]{revtex}

\tighten
\preprint {Version 0.85 - 12.9.98}
\newcommand{\be}{\begin{equation}}
\newcommand{\ee}{\end{equation}}

\begin{document}
\title{High-field transport properties of bulk Si: A test for the Fokker-Planck
approach}
\author{F. Comas* and Nelson Studart}
\address{Departmento de F\'{i}sica, Universidade Federal de S\~{a}o Carlos,\\
13565-905, S\~{a}o Carlos, S\~{a}o Paulo, Brazil}
\maketitle

\begin{abstract}
High electric-field transport parameters are calculated using an analytical
Fokker-Planck approach (FPA), where transport is modeled as a
drift-diffusion process in energy space. We have applied the theory to the
case of Si, taking into account the six intervalley phonons, aiming to test
the FPA. The obtained results show a quite reasonable agreement with
experimental data and Monte Carlo simulations confirming in this case that
the FPA works very well for high enough electric fields.

PACS numbers: 72.10.Di; 72.20.Ht.
\end{abstract}

\vspace{0.9cm}%

More than three decades ago the Fokker-Planck approach (FPA) was proposed as
an alternative for the Boltzmann transport equation (BTE) in the calculation
of semiconductor transport properties \cite{b1,b2}. Recently the theory has
been revisited in a series of works \cite{b3,b4,b5,b6,b7} where a relatively
detailed analysis of both mathematical and physical aspects of this
formalism was developed. In these papers the same model system was
considered in calculations using the FPA and the BTE by means of the Monte
Carlo method. The results of both approaches showed good agreement in the
high electric-field regime for the mentioned model system. In spite of this,
there remains a certain degree of doubt about how the FPA could handle a
more realistic model of a concrete semiconductor with several possible
scattering mechanisms and complicated band structure. The FPA considers
transport, in the opposite regime of the ballistic one, as a certain
diffusive-drifting ``motion'' of the carriers in the energy space and it is
valid when $\tau (\vec{p})<<t<<\tau _{E},$ where $\tau (\vec{p})$ is the
momentum relaxation time and $\tau _{E}$ is the energy relaxation time. The
method is semiclassical by its own nature and applicable when the energy
exchange between the carriers and the surrounding medium can be assumed
quasicontinuous which excludes highly inelastic scattering processess. It is
valid when the average carrier energy is much larger than the exchanged
energy as in the case of high-field transport. The FPA has the advantage of
being analytical, and, whenever it can be applied, saves computational time
and allows a more transparent physical interpretation.

In this report, we present results within the FPA for bulk Si and compare
them with experimental data and previous Monte Carlo simulations (using the
BTE). We show that the FPA leads to good results as compared with those data
whatever several scattering mechanisms (six intervalley phonons between the
Si $\Delta $ valleys in the conduction band) are taken into account. The
intravalley acoustic phonons were ignored, and thus our results are reliable
just for high enough temperatures.

The evolution of the energy distribution function (DF) $f(E,t)$ is governed
by the Fokker-Planck equations\cite{b1,b2,b3}

\begin{equation}
\frac{\partial }{\partial t}f(E,t)+\frac{1}{N(E)}\frac{\partial }{\partial E}%
J(E,t)=0,  \label{e1}
\end{equation}
where

\begin{equation}
J(E,t)=W(E)N(E)f(E,t)-\frac{\partial }{\partial E}\left[
D(E)N(E)f(E,t)\right] ,  \label{e2}
\end{equation}
such that $f(E,t)N(E)$ gives the number of carriers at time $t$ with
energies in the interval $(E,E+dE)$, while the function $N(E)$ represents
the density of states (DOS). In Eq. (\ref{e2}) $W(E)$ represents a certain
``drift velocity'' in energy space and in fact gives the rate of energy
balance of the carrier, $D(E)$ is a kind of diffusion coefficient and $%
J(E,t) $ represents thus the carrier current density in energy space.\cite
{b1,b2,b3} Under steady state conditions, the Eq. (\ref{e2}) transforms into

\begin{equation}
\frac{\partial }{\partial E}\left[ D(E)N(E)f(E)\right] =W(E)N(E)f(E).
\label{e3}
\end{equation}

The carriers interact with the phonons and the applied dc electric field $%
\vec{F}$. We assume a phonon reservoir in thermal equilibrium at the
temperature $T$ and that the continuous exchange of phonons between the
carriers and the bath does not affect the thermal equilibrium of the latter.
Hence, the coefficients $W(E)$ and $D(E)$ are split as follows

\begin{equation}
D(E)=D_{F}(E)+D_{ph}(E)\quad ,\quad W(E)=W_{F}(E)+W_{ph}(E).  \label{e4}
\end{equation}
The label ''$F$'' (''$ph$'') denotes the electric field (phonon)
contribution to these coefficients, whose explicit forms will be given
below. Equation (\ref{e3}) has the simple solution

\begin{equation}
f(E)=\exp \left\{ \int \left[ \frac{W_{ph}(E)}{D_{F}(E)}dE\right] \right\} ,
\label{e5}
\end{equation}
where $D_{ph}(E)$ was neglected. This approximation is very well fulfilled
in all the cases of interest for us \cite{b3}.

We consider transport of electrons in the Si conduction band (CB) in a high
dc electric field regime ($F>10$ kV/cm). We take into account the six
ellipsoidal energy valleys of Si at the $\Delta $ points of the Brillouin
zone (along the $<100>$ direction). To be specific, let us take $\vec{F}%
=(0,0,F)$, where the $z$ axis is taken along one high symmetry direction,
and denote by ``$l$'' (``$tr$'') the valleys with principal axis parallel
(perpendicular) to $\vec{F}$.

The explicit expressions for $D_{F}(E)$ and $W_{ph}(E)$ are\cite{b2,b3}

\begin{equation}
D_{F}(E)=\left\langle \tau (\vec{p})[q\vec{F}\cdot \nabla _{\vec{p}}\epsilon
(\vec{p})]^{2}\right\rangle ,  \label{e8}
\end{equation}

\begin{equation}
W_{ph}(E)=\hbar \omega \left[ 1/\tau _{abs}(\vec{p})-1/\tau _{em}(\vec{p}%
)\right] ,  \label{e9}
\end{equation}
where the brackets represent an average over the constant energy surface $%
\epsilon (\vec{p})=const$, $\tau (\vec{p})$ is the {\it total} relaxation
time due to the electron-phonon scattering, and ``{\it abs}'' (``{\it em}'')
denotes phonon absorption (emission) by the electron due to the several
scattering mechanisms.

A straightforward evaluation of Eq.(\ref{e8}) leads to

\begin{equation}
D_{F}^{j}(E)=\frac{2e^{2}F^{2}}{3m_{j}}\tau (\epsilon )\gamma (\epsilon
)/(\gamma ^{\prime }(\epsilon ))^{2}\quad j=l,tr,  \label{e11}
\end{equation}
where $\gamma ^{\prime }(\epsilon )$ denotes the derivative of the function $%
\gamma (\epsilon )=\epsilon (1+\alpha \epsilon )$ responsible by the
non-parabolicity of the band structure with $\epsilon =\epsilon (\vec{p})$
being the energy dispersion for a given valley. For $\tau (\epsilon )$ we
shall consider the six intervalley phonons responsible for transitions
between the equivalent $\Delta $-valleys of Si. Then the total relaxation
time reads as

\[
\frac{1}{\tau (\epsilon )}=\sum_{i=1}^{6}C_{oi}\left[ n_{i}(T)\sqrt{\gamma
(\epsilon +\hbar \omega _{i})}|1+2\alpha (\epsilon +\hbar \omega
_{i})|\right. 
\]
\begin{equation}
\left. +(n_{i}(T)+1)\sqrt{\gamma (\epsilon -\hbar \omega _{i})}|1+2\alpha
(\epsilon -\hbar \omega _{i})|\Theta (\epsilon -\hbar \omega _{i})\right] ,
\label{e12}
\end{equation}
where $\Theta (\epsilon )$ is the step function, $n_{i}(T)$ is the phonon
distribution function, and $C_{oi}=(m_{tr}m_{l}^{1/2}D_{oi}^{2})/(\sqrt{2}%
\pi \rho \hbar ^{3}\omega _{i})$ with $m_{tr}$ ($m_{l}$) being the
transverse (longitudinal) effective mass, $\rho $ the semiconductor density, 
$\omega _{i}$ and $D_{oi}$ are the frequency and deformation-potential
constant respectively for intervalley phonons of type $i$ \cite{b9}. In Eq.(%
\ref{e12}), the first and second terms correspond to $1/\tau
_{abs}^{i}(\epsilon )$ and $1/\tau _{em}^{i}(\epsilon )$ respectively. The
phonon contribution can be written as

\begin{equation}
W_{ph}(\epsilon )=\sum_{i=1}^{6}W_{ph}^{i}(\epsilon ).  \label{e17}
\end{equation}
where $W_{ph}^{i}(\epsilon )$ is given by Eq.(\ref{e9}) for each $i.$
Intravalley optical phonons do not contribute to transition rates, because
the corresponding transitions are forbidden by the selection rules and we
assume that, for high $T$ and $E,$ the contribution of intravalley acoustic
phonons should be neglected.

Once, we have evaluated $W_{ph}(E)$ and $D_{F}^{j}(E)$, we have to calculate
the integral in Eq.(\ref{e5}) to obtain the two DF, $f_{l}(E)$ and $%
f_{tr}(E) $, corresponding to the $l$-$tr$ valleys respectively. Of course,
the FPA is of practical use only in the case when such integration can be
analytically performed which is not the case for expressions of $D_{F}^{j} $
and $1/\tau (E)$ given by Eqs.(\ref{e11}) and (\ref{e12}). So, we consider a
single effective intervalley phonon with energy $\hbar \omega _{0}=0.0343\, $
eV, obtained from an average of different phonon frequencies given in Table
VI of \cite{b9}, and with constant $D_{o}$ obtained by the superposition of
the different deformation-potential constants, but also including the number
of final equivalent valleys for each kind of transition. With this
approximation we obtain the following result

\begin{equation}
\sum_{i=1}^{6}\frac{W_{ph}^{i}(E)}{D_{F}^{j}(E)}=\frac{3m_{j}\hbar \omega
_{0}}{2e^{2}F^{2}}\sum_{i,k}C_{oi}C_{ok}\Phi (E,T),\quad j=l,tr,  \label{e18}
\end{equation}
with

\[
\Phi (E,T)=n^{2}(T)(E+1)(1+E_{0}(E+1))(1+2E_{0}(E+1))^{2} 
\]
\begin{equation}
-(n(T)+1)^{2}(E-1)(1+E_{0}(E-1))(1+2E_{0}(E-1))^{2}\Theta (E-1).  \label{e19}
\end{equation}
where hereafter $E$ is in units of $\hbar \omega _{0}$ and $E_{0}=\hbar
\omega _{0}\alpha $. Considering the contributions from different valleys
and using Eq.(\ref{e5}), again applying the parameters from Table VI Ref.%
\cite{b9}, we are led to:

\begin{equation}
f(E)=\exp \left[ \beta _{j}\int \Phi (x,T)dx\right] ,  \label{e20}
\end{equation}
with $\beta _{j}=(3m_{j}m_{d}^{3}D_{o}^{4})/4\pi ^{2}\rho
^{2}e^{2}F^{2}\hbar ^{4}$ and $D_{o}$ is the effective deformation-potential
constant defined through $D_{o}^{4}=\sum D_{oi}^{2}D_{ok}^{2}$. We have
estimated $D_{o}=12.09\times 10^{8}$ eV/cm and in all the above expressions
the overlapping integral (see Ref.\cite{b9}) was taken equal to unity.
However, for numerical computations it should be useful consider it as a
fitting parameter.

The integral involved in Eq.(\ref{e20}) can be analytically performed in a
straightforward way and the general structure of the DF has the form:

\begin{equation}
f(E)=E^{A}(1+E_{0}E)^{B}\exp (C\cdot P(E)),  \label{e22}
\end{equation}
where $A$, $B$ and $C$ are parameters dependent on $T$ and $F$ and $P(E$) is
a polynomial.. This structure is far from a Maxwellian one. The DF describes
the stationary non-equilibrium configuration where an electron temperature $%
T_{e}$ cannot be defined. From Eq.(\ref{e22}), we can immediately obtain the
average electron energy, estimated as $E_{av}(T,F)=(E_{av}^{l}+2
E_{av}^{tr})/3$ with

\begin{equation}
E_{av}^{j}=\left[ \int Ef_{j}(E)N(E)dE\right] /\int f(E)_{j}N(E)dE\quad
j=l,\,tr.  \label{e23}
\end{equation}

In Fig. 1, the electric-field dependence of $E_{av}$ is depicted for
different temperatures. As expected, we found a weak temperature dependence.
We can see that the average electron energy increases for larger electric
fields. Moreover, the condition $E_{av}>>\hbar \omega _{0}$ is fairly well
accomplished, ensuring that the FPA is within its range of validity for the
given temperatures. Our results cannot be expected to be correct for low
temperatures (or too low carrier energies) because we neglect intravalley
acoustic phonons.

The drift velocity $v_{d}=(v_{dl}+2v_{dtr})/3$ can be also evaluated from%
\cite{b2,b3}

\begin{equation}
v_{dj}=\frac{2eF}{3m_{j}}\int \frac{\gamma (E)\tau (E)}{(\gamma ^{\prime
}(E))^{2}}\left[ -\frac{df_{j}(E)}{dE}\right] N(E)dE/\int f(E)_{j}N(E)dE.
\label{e25}
\end{equation}
In Fig. 2, we show $v_{d}$ as a function of $F$ for two different
temperatures. We see that the general behavior of the curve is qualitatively
correct if compared with experimental results and Monte Carlo simulations.
For a more quantitative comparison, we present in Fig. 3 our results
together with the experimental data and those from Monte Carlo simulations%
\cite{b10,b11}. As it can be seen, we obtained a good agreement with both
results for $T=300$ K in the high electric-field regime. However, in the
opposite limit, the results from FPA do not reproduce those from experiments
and Monte Carlo calculations, as should be expected. For comparison with
experimental data we have taken $\beta _{j}$ as a fitting parameter, an
issue which can be reasonably understood considering the overlapping
integral for the electron-phonon scattering probabilities. Another point to
be stressed is that reliable results were obtained just for high
temperatures. However this is not a limitation of the FPA itself but of our
present calculations since we have ignored intravalley acoustic phonons.

In conclusion, we have shown that transport problems can be tackled by the
mathematically simple FPA even in the case of a concrete semiconductor. We
can also emphasize that the possibility for achieving correct results from
the FPA depends on the chance of performing good enough approximations for
the relaxation processes, allowing the analytical evaluation of the integral
in Eq.(\ref{e5}) and still retaining the essential physical picture. The
results, being acceptable just for the higher electric fields, in fact are
very close to those of Monte-Carlo simulations. Additional calculations for
T= 430 K also revealed good agreement with those of \cite{b10}. The
saturation effect, however, is not predicted by FPA. For higher electric
field intensities a slow but ever decreasing behaviour is achieved.

We acknowledge financial support from the Funda\c{c}\~{a}o de Amparo \`{a}
Pesquisa de S\~{a}o Paulo. F. C. is grateful to Departamento de F\'{i}sica,
Universidade Federal de S\~{a}o Carlos, for hospitality.

\medskip 

\noindent *Permanent Address: Depto. de F\'{i}sica Te\'{o}rica, Universidad.
de la Habana, Vedado 10400, Havana, Cuba.

\medskip 

\begin{figure}[tbp]
\caption{Average electron energy (in units of $\hbar \omega_0$) as a
function of the electric field for three different temperatures. It is
clearly seen that the average energy is actually much larger than the phonon
energy.}
\label{1}
\end{figure}
\begin{figure}[tbp]
\caption{Drift velocity for three different temperatures as a function of
the electric field. The general trend of the curves corresponds with what is
seen in experiments and Monte Carlo simulations.}
\label{2}
\end{figure}
\begin{figure}[tbp]
\caption{Drift velocity as a function of electric field for 300 K for Si.
Our results (solid curve) are compared with Monte Carlo simulations (dotted
curve) and experimental data . }
\label{3}
\end{figure}

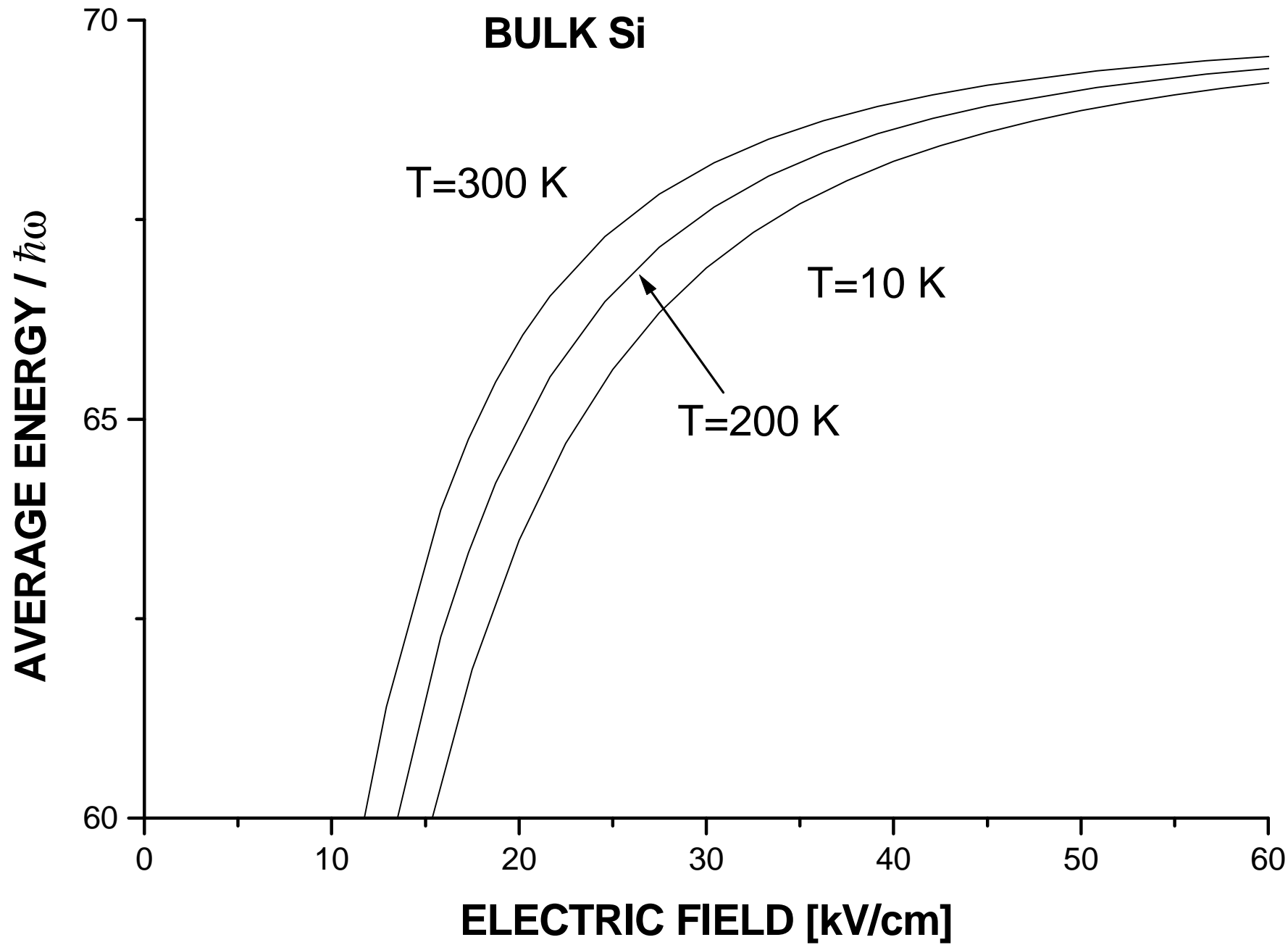
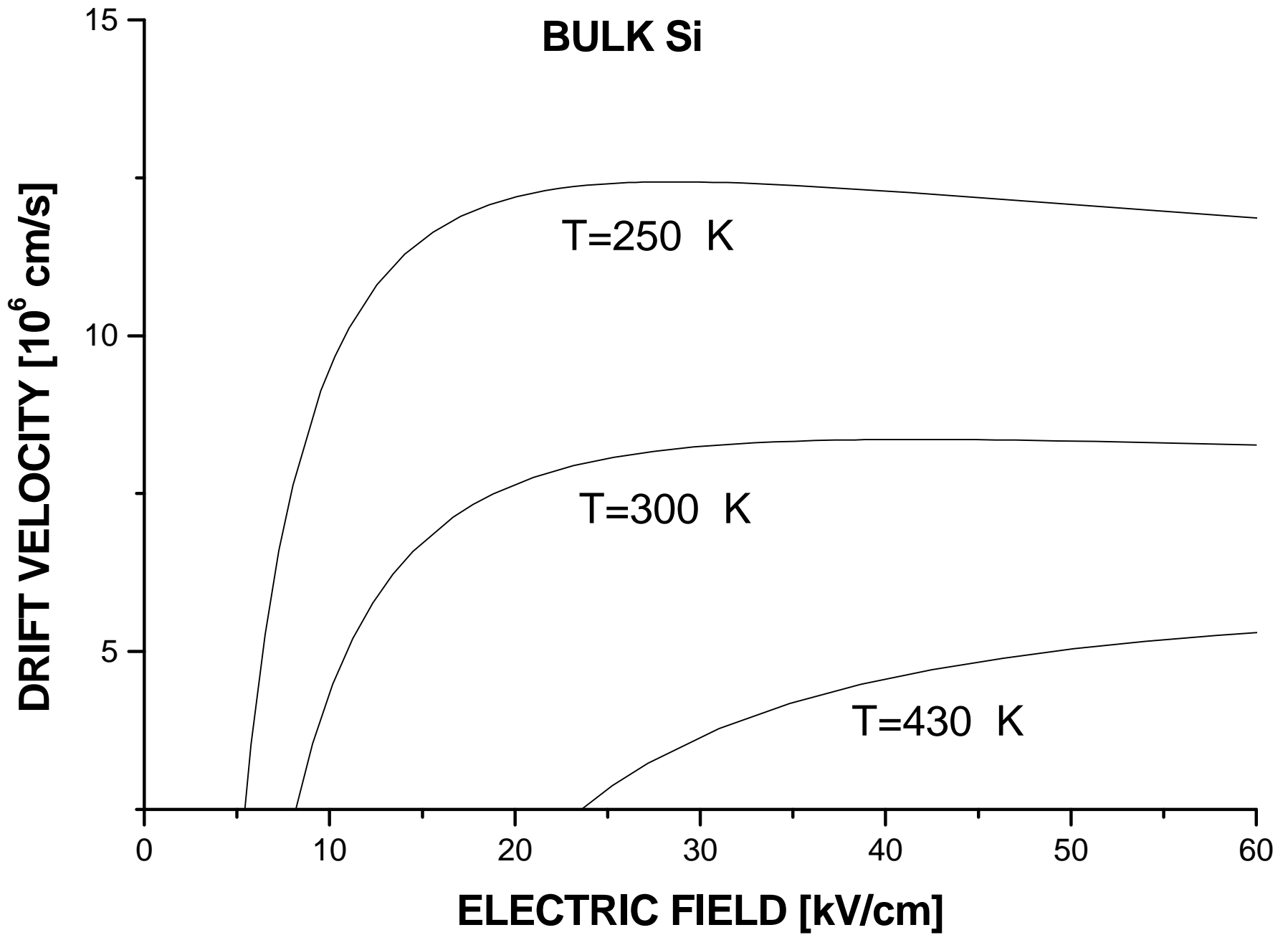
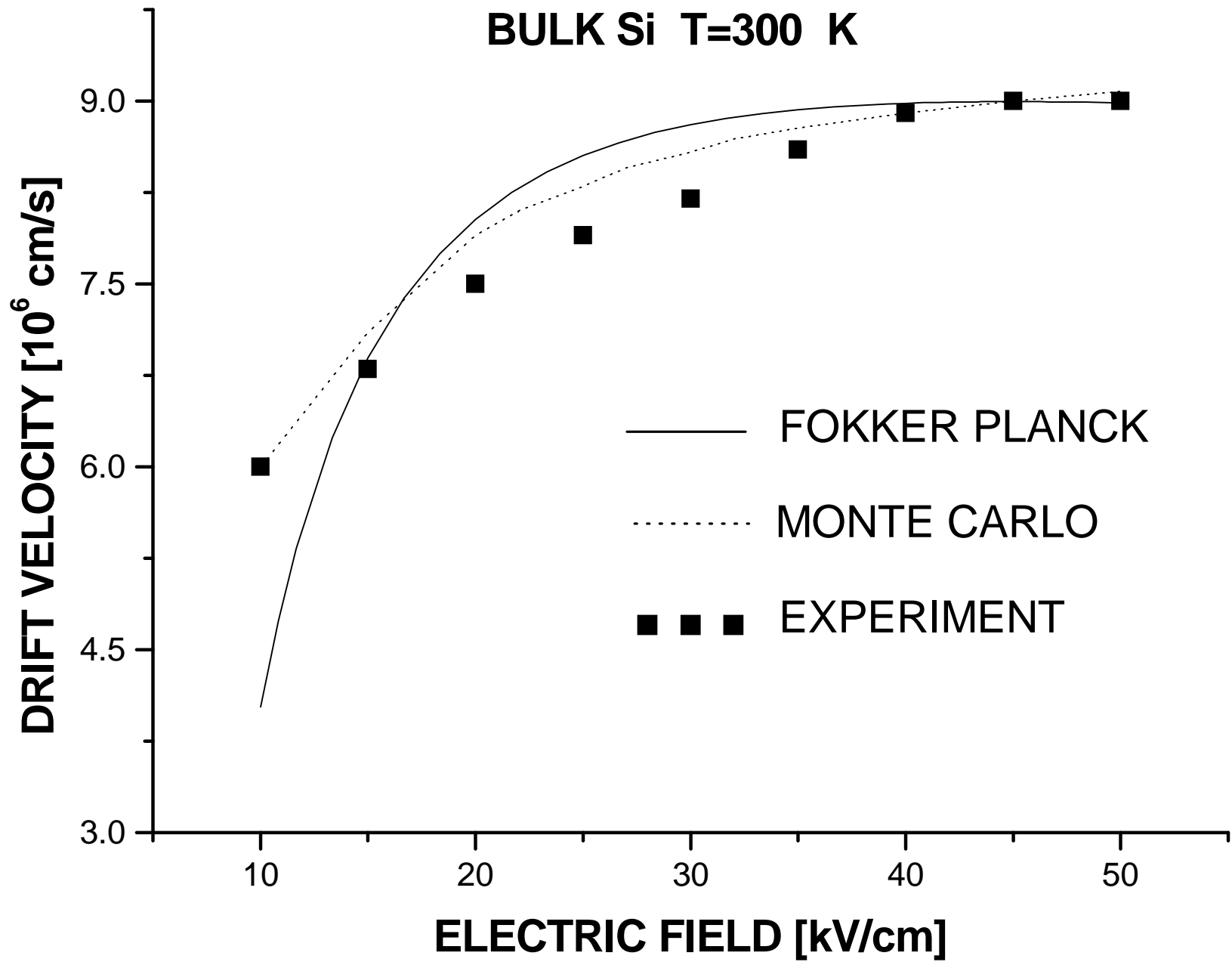
\end{document}